\documentclass[preprint,aps,12pt,showpacs,nofootinbib,tightenlines]{revtex4}
\usepackage{amsmath}
\usepackage{amssymb}
\usepackage{epsfig}
\usepackage{graphicx}
\newcommand{\psl}{ P \hspace{-2.4truemm}/}
\newcommand{\epsl}{\epsilon \hspace{-1.8truemm}/\,  }
\def\be{\begin{eqnarray}}
\def\en{\end{eqnarray}}
\def\non{\nonumber\\}

\def\ra{\rangle}

\def\sl{\!\!\!\slash}
\def\prd{{Phys. Rev. D}~}
\def\prl{{ Phys. Rev. Lett.}~}
\def\plb{{ Phys. Lett. B}~}

\def\epjc{{ Eur. Phys. J. C}~}
\newcommand{\acp}{{\cal A}_{CP}}
\begin{document}
\title{Study of scalar meson $f_0(980)$ and $K_0^*(1430)$ from $B \to f_0(980)\rho(\omega, \phi)$ and $B \to K^*_0(1430)\rho(\omega)$ Decays }
\author{Zhi-Qing Zhang
\footnote{Electronic address: zhangzhiqing@haut.edu.cn} } 
\affiliation{\it \small  Department of Physics, Henan University of
Technology, Zhengzhou, Henan 450052, P.R.China } 
\date{\today}
\begin{abstract}
In the two-quark model
supposition for $f_0(980)$ and $K_0^{*}(1430)$, the branching ratios and the direct CP-violating
asymmetries for decays $\bar{B}^0\to f_0(980)\rho^0(\omega,\phi),
K^{*0}_0(1430)\rho^0(\omega),K^{*-}_0(1430)\rho^+$ and $B^-\to f_0(980)\rho^-, K^{*0}_0(1430)\rho^-, K^{*-}_0(1430)\rho^0(\omega)$
are studied by employing the perturbative QCD (PQCD)factorization approach. we find the following results: (a) if the scalar meson $f_0(980)$
is viewed as a mixture of $s\bar s$ and $(u\bar u+d\bar d)/\sqrt{2}$, the branching ratios of the $b\to d$ transition processes
$\bar{B}^0\to f_0(980)\rho^0(\omega,\phi)$ and $B^-\to f_0(980)\rho^-$ are smaller than the
currently experimental upper limits, and
the predictions for the decays $\bar{B}^0\to f_0(980)\omega, B^-\to f_0(980)\rho^-$ are not far away from their limits; (b) in
the $b\to s$ transition processes $B\to K^*_0(1430)\rho(\omega)$, the branching ratio of $\bar B^0\to K^{*0}_0(1430)\rho^0$
is the smallest one, at the order of $10^{-7}$ by treating $K^*_0(1430)$ as the lowest lying state, about $4.8\times10^{-6}$ by considering
$K^*_0(1430)$ as the first excited state; (c) the
direct CP-violating asymmetries of decays $B\to f_0(980)\rho(\omega)$ have a strong dependence on the mixing angle $\theta$: they are large
in the range of $25^\circ<\theta<40^\circ$, and small in the range of $140^\circ<\theta<165^\circ$, while the
direct CP-violating asymmetry amplitudes of decays $B\to K^{*}_0(1430)\rho(\omega)$ are not large in the two kinds of state supposition
for $K^*_0(1430)$ and most of them are less than $20\%$.
\end{abstract}

\vspace{1cm}

\maketitle


\section{Introduction}\label{intro}
Along with many scalar mesons are found in experiments, more and more efforts have been made to study the scalar meson spectrum
theoretically \cite{nato,jaffe,jwei,baru,celenza,stro,close1}. Today, it is still a difficult but interesting topic. Our most important
task is to uncover the
mysterious structure of the scalar mesons.  There are two typical schemes for the classification to
them \cite{nato,jaffe}. The scenario I (SI): the nonet mesons below
1 GeV, including $f_0(600), f_0(980), K^*(800)$ and $a_0(980)$, are
usually viewed as the lowest lying $q\bar q$ states, while the nonet
ones near 1.5 GeV, including $f_0(1370), f_0(1500)/f_0(1700),
K^*(1430)$ and $a_0(1450)$, are suggested as the first excited
states. In the scenario II (SII), the nonet mesons near 1.5 GeV are
treated as $q\bar q$ ground states, while the nonet mesons below 1
GeV are exotic states beyond the quark model such as four-quark
bound states.

The production of the scalar mesons from B meson decays provides a different unique insight to the inner structures of these mesons.
It provides various factorization approaches a new usefulness. The QCD fractorization (QCDF) approach \cite{ccysp,ccysv}have been used to
systematically
study the $B$ meson decays with a scalar meson involved in the final states. The authors draw the conclusion that scenario II is more
preferable than scenario I, that is to say, the light scalar mesons below 1 GeV are possible four-quark bound states and the scalar mesons near 1.5 GeV
are the lowest lying $q\bar q$ states. If $f_0(980)$ is a four-quark bound state, it requires to pick up two energetic quark-antiquark
pairs to form this scalar meson,
so one expects that the $B\to f_0(980)X$ rate might be smaller in the four-quark model than in the two-quark picture.
From the previous
calculations \cite{wwang,ccysp}, we also expect that the two-quark component of $f_0(980)$ plays an essential role for
$B\to f_0(980)\rho(\omega,\phi)$ decays. Just like QCDF approach, in order to make quantitative
prediction, we assume the scalar meson $f_0(980)$ as a mixture of $s\bar{s}$ and $n\bar{n}(\equiv(u\bar{u}+d\bar{d})/\sqrt{2})$, that is
\be
|f_0(980)\ra = |s\bar s\ra\cos\theta+|n\bar n\ra\sin\theta,
\en
where $\theta$ is the $f_0-\sigma$ mixing angle. In
the phenomenal and experimental analyses \cite{theta,theta1}, $\theta$ lies in the ranges of $25^\circ<\theta<40^\circ$ and
$140^\circ<\theta<165^\circ$. Certainly, $K^*_0(1430)$
can be treated as a $q\bar q$ state in both SI and  SII, so we will calculate $B\to K^*_0(1430)\rho(\omega)$ decays in two scenarios.

On the experimental side, for $f_0(980)$ emerging as a pole of the amplitude in the S wave \cite{kami}, many
channels such as $B\to f_0(980)K$ can be obtained by fitting  of Dalitz plots of the decays $B\to \pi^+\pi^-K$ and
$B\to \bar KKK$ and so on\cite{Bellef01,Bellef02,BaBarf01,BaBarf02}. Although many such
decay channels that involved $f_0(980)$ in the final states have
been measured over the years, it has yet not been possible to account for its inner structure.
For the decays $B\to f_0(980)\rho(\omega,\phi)$,  only the upper limits are available now \cite{pdg08,barbar0}:
\be
Br(B^-\to f_0(980)\rho^-) < 3.8\times 10^{-6},\non
Br(\bar B^0\to f_0(980)\rho^0) < 1.06\times 10^{-6}, \non
Br(\bar B^0\to f_0(980)\omega) < 3.0\times 10^{-6},\non
Br(\bar B^0\to f_0(980)\phi) < 7.6\times 10^{-7}.
\en
It is noticed that we have assumed $Br(f_0(980)\to\pi^+\pi^-)=0.50$ to obtain the upper data.
For the decays $B\to K^*_0(1430)\rho(\omega)$, there is still no experimental result.

Here we would like to use the perturbative QCD (PQCD) approach to study $f_0(980)$ and $K_0^*(1430)$
in the decays $B \to f_0(980)\rho(\omega, \phi)$ and $B \to K^*_0(1430)\rho(\omega)$.
In the following, $f_0(980)$ and $K^*_0(1430)$ are denoted as $f_0$ and $K^*_0$ in some places for convenience.
The layout of this paper is as follows. In section \ref{proper}, the relevant decay constants
and light-cone distribution amplitudes of relevant mesons are introduced.
In section \ref{results}, we then analysis these decay channels using the PQCD approach.
The numerical results and the discussions are given
in section \ref{numer}. The conclusions are presented in the final part.


\section{decay constants and distribution amplitudes }\label{proper}

For the wave function of the heavy B meson,
we take:
\be
\Phi_B(x,b)=
\frac{1}{\sqrt{2N_c}} (\psl_B +m_B) \gamma_5 \phi_B (x,b).
\label{bmeson}
\en
Here only the contribution of Lorentz structure $\phi_B (x,b)$ is taken into account, since the contribution
of the second Lorentz structure $\bar \phi_B$ is numerically small \cite{cdlu} and has been neglected. For the
distribution amplitude $\phi_B(x,b)$ in Eq.(\ref{bmeson}), we adopt the model
\be
\phi_B(x,b)=N_Bx^2(1-x)^2\exp[-\frac{M^2_Bx^2}{2\omega^2_b}-\frac{1}{2}(\omega_bb)^2],
\en
where $\omega_b$ is a free parameter, and the value of the normalization factor is taken as $N_B=91.745$ for $\omega_b=0.4$
in numerical calculations.

In  two-quark picture, the vector decay constant $f_{S}$ and the scalar decay constant
$\bar {f}_{S}$ for a scalar meson $S$ can  be defined as:
\be
\langle S(p)|\bar q_2\gamma_\mu q_1|0\ra&=&f_{S}p_\mu,
\en
\be
\langle S(p)|\bar q_2q_1|0\ra=m_{S}\bar {f}_{S}, \label{fbar}
\en
where $m_S(p)$ is the mass (momentum) of the scalar meson. The relation between $f_S$ and
$\bar f_S$ is
\be
\frac{m_S}{m_2(\mu)-m_1(\mu)}f_S=\bar f_S,
\en
where $m_{1,2}$ are the running current quark masses.
For the neutral scalar meson $f_0$, owing to charge conjugation invariance or the G parity conservation, it
cannot be produced via the vector current, so $f_{f_0}=0$. Taking the mixing into account,
Eq.(\ref{fbar}) is changed to:
\be \langle f_0^n|\bar
dd|0\ra=\langle f_0^n|\bar uu|0\ra=\frac{1}{\sqrt 2}m_{f_0}\tilde
f^n_{f_0},\,\,\,\, \langle f_0^n|\bar ss|0\ra=m_{f_0}\tilde
f^s_{f_0}.
\en
Because the decay constants $\tilde f_{f_0}^n$ and $\tilde f_{f_0}^s$
are very close\cite{ccysp}, we assume that $\tilde
f_{f_0}^n=\tilde f_{f_0}^s$ and denote them as $\bar f_{f_0}$ in the
following. For the scalar meson $K^*_0(1430)$, $f_{K^*_0}$ will get a very small value
after the $SU(3)$ symmetry breaking being considered.
The light-cone distribution amplitudes (LCDAs) for the  scalar
meson $S$ can be written as:
\be \langle
S(p)|\bar q_1(z)_l q_2(0)_j|0\rangle
&=&\frac{1}{\sqrt{2N_c}}\int^1_0dx \; e^{ixp\cdot z}\non
&&
\times \{ p\sl\Phi_{S}(x)
+m_{S}\Phi^S_{S}(x)+m_{S}(n\sl_+n\sl_--1)\Phi^{T}_{S}(x)\}_{jl},\quad\quad\label{LCDA}
\en
here $n_+$ and $n_-$ are light-like vectors:
$n_+=(1,0,0_T),n_-=(0,1,0_T)$, and $n_+$ is parallel with the moving direction of the scalar meson.
The normalization can be related to
the decay constants:
\be \int^1_0 dx\Phi_{S}(x)=\int^1_0
dx\Phi^{T}_{S}(x)=0,\,\,\,\,\,\,\,\int^1_0
dx\Phi^{S}_{S}(x)=\frac{\bar f_{S}}{2\sqrt{2N_c}}\;.
\en
The twist-2 LCDA $\Phi_S$ can be expanded in the Gegenbauer polynomials:
\be
\Phi_{S}(x,\mu)&=&\frac{\bar f_{S}(\mu)}{2\sqrt{2N_c}}6x(1-x)\left[B_0(\mu)+\sum_{m=1}^\infty B_m(\mu)C^{3/2}_m(2x-1)\right],
\en
where the decay constants and the Gegenbauer moments $B_1,B_3$ of distribution amplitudes for $f_0(980)$ and $K^*(1430)$ have been
calculated in the QCD sum rules\cite{ccysp}. These values are all scale dependent and specified below:
\be
{\rm scenario I:} B_1(K^*_0)&=&0.58\pm0.07, B_3(K^*_0)=-1.2\pm0.08, \bar f_{K^*_0}=-(300\pm30){\rm MeV},\\
                  B_1(f_0)&=&-0.78\pm0.08, B_3(f_0)=0.02\pm0.07, \bar f_{f_0}=-(370\pm20){\rm MeV};\\
{\rm scenario II:}B_1(K^*_0)&=&-0.57\pm0.13, B_3(K^*_0)=0.42\pm0.22, \bar f_{K^*_0}=-(445\pm50){\rm MeV},\quad
\en
which are taken by fixing the scale at 1GeV.

As for the twist-3 distribution amplitudes $\Phi_{S}^S$ and $\Phi_{S}^T$, we adopt the asymptotic form:
\be
\Phi^S_{S}&=& \frac{1}{2\sqrt {2N_c}}\bar f_{S},\,\,\,\,\,\,\,\Phi_{S}^T=
\frac{1}{2\sqrt {2N_c}}\bar f_{S}(1-2x).
\en

The distribution amplitudes up to twist-3 of the vector mesons are
\be
\langle V(P,\epsilon^*_L)|\bar q_{2\beta}(z)q_{1\alpha}(0)|0\rangle=\frac{1}{2N_C}\int^1_0dxe^{ixP\cdot z}[M_V\epsl^*_L\Phi_V(x)
+\epsl_L^*\psl\Phi_V^t(x)+M_V\Phi^s_V(x)]_{\alpha\beta},\quad
\en
for longitudinal polarization. The distribution amplitudes
can be parametrized as
\be
\Phi_V(x)&=&\frac{2f_V}{\sqrt{2N_C}}[1+a^{\|}_2C^{\frac{3}{2}}_2(2x-1)],\\
\Phi_V^t(x)&=&\frac{3f^T_V}{2\sqrt{2N_C}}(2x-1)^2,\quad \phi_V^s(x)=-\frac{3f^T_V}{2\sqrt{2N_C}}(2x-1),
\en
where the decay constant $f_V$ \cite{yao} and the transverse decay constant $f^T_V$ \cite{pball} are given as the following values:
\be
f_\rho&=&209\pm2 {\rm MeV}, f_\omega=195\pm3 {\rm MeV}, f_\phi=231\pm4 {\rm MeV}, \\
f^T_\rho&=&165\pm9 {\rm MeV}, f^T_\omega=151\pm9 {\rm MeV}, f^T_\phi=186\pm9 {\rm MeV}.
\en
Here the Gegenbauer polynomial is defined as $C^{\frac{3}{2}}_2(t)=\frac{3}{2}(5t^2-1)$. For the Gegenbauer moments, we quote the
numerical results as \cite{pball1}:
\be
a^{\|}_{2\rho}=a^{\|}_{2\omega}=0.15\pm0.07, a^{\|}_{2\phi}=0.18\pm0.08.
\en


\section{ the perturbative QCD  calculation} \label{results}

Under the two-quark model for the scalar mesons $f_0$ and $K^*_0$ supposition,
the decay amplitude for $B\to VS$, where $V$ represents $\rho, \omega, \phi$ and $S$ represents $f_{0},K^*_0$,
 can be conceptually written as the convolution,
\be
{\cal A}(B \to V S)\sim \int\!\! d^4k_1
d^4k_2 d^4k_3\ \mathrm{Tr} \left [ C(t) \Phi_B(k_1) \Phi_{V}(k_2)
\Phi_{S}(k_3) H(k_1,k_2,k_3, t) \right ], \label{eq:con1}
\en
where $k_i$'s are momenta of the anti-quarks included in each mesons, and
$\mathrm{Tr}$ denotes the trace over Dirac and color indices. $C(t)$
is the Wilson coefficient which results from the radiative
corrections at short distance. In the above convolution, $C(t)$
includes the harder dynamics at larger scale than $M_B$ scale and
describes the evolution of local $4$-Fermi operators from $m_W$ (the
$W$ boson mass) down to $t\sim\mathcal{O}(\sqrt{\bar{\Lambda} M_B})$
scale, where $\bar{\Lambda}\equiv M_B -m_b$. The function
$H(k_1,k_2,k_3,t)$ describes the four quark operator and the
spectator quark connected by
 a hard gluon whose $q^2$ is in the order
of $\bar{\Lambda} M_B$, and includes the
$\mathcal{O}(\sqrt{\bar{\Lambda} M_B})$ hard dynamics. Therefore,
this hard part $H$ can be perturbatively calculated. The function
$\Phi_{(V, S)}$ are the wave functions of the vector meson $V$ and
the scalar meson $S$, respectively.

Since the $b$ quark is rather heavy we consider the $B$ meson at rest
for simplicity. It is convenient to use light-cone coordinate $(p^+,
p^-, {\bf p}_T)$ to describe the meson's momenta, \be p^\pm =
\frac{1}{\sqrt{2}} (p^0 \pm p^3), \quad {\rm and} \quad {\bf p}_T =
(p^1, p^2). \en Using these coordinates the $B$ meson and the two
final state meson momenta can be written as \be P_B =
\frac{M_B}{\sqrt{2}} (1,1,{\bf 0}_T), \quad P_{2} =
\frac{M_B}{\sqrt{2}}(1-r^2_S,r^2_V,{\bf 0}_T), \quad P_{3} =
\frac{M_B}{\sqrt{2}} (r^2_S,1-r^2_V,{\bf 0}_T), \en respectively, where the ratio $r_{S(V)}=m_{S(V)}/M_B$, and
$m_{S(V)}$ is the scalar meson S (the vector meson V) mass. Putting the anti-quark momenta in $B$,
$V$ and $S$ mesons as $k_1$, $k_2$, and $k_3$, respectively, we can
choose
\be k_1 = (x_1 P_1^+,0,{\bf k}_{1T}), \quad k_2 = (x_2
P_2^+,0,{\bf k}_{2T}), \quad k_3 = (0, x_3 P_3^-,{\bf k}_{3T}). \en
For these considered decay channels, the integration over $k_1^-$,
$k_2^-$, and $k_3^+$ in eq.(\ref{eq:con1}) will lead to
\be
 {\cal
A}(B \to V S) &\sim &\int\!\! d x_1 d x_2 d x_3 b_1 d b_1 b_2 d
b_2 b_3 d b_3 \non && \cdot \mathrm{Tr} \left [ C(t) \Phi_B(x_1,b_1)
\Phi_{V}(x_2,b_2) \Phi_{S}(x_3, b_3) H(x_i, b_i, t) S_t(x_i)\,
e^{-S(t)} \right ], \quad \label{eq:a2}
\en
where $b_i$ is the
conjugate space coordinate of $k_{iT}$, and $t$ is the largest
energy scale in function $H(x_i,b_i,t)$.
In order to smear the end-point singularity on $x_i$,
the jet function $S_t(x)$ \cite{li02}, which comes from the
resummation of the double logarithms $\ln^2x_i$, is used.
The last term $e^{-S(t)}$ in Eq.(\ref{eq:a2}) is the Sudakov form factor which suppresses
the soft dynamics effectively \cite{soft}.

 For the considered decays, the related weak effective
Hamiltonian $H_{eff}$ can be written as \cite{buras96}
\be
\label{eq:heff} {\cal H}_{eff} = \frac{G_{F}} {\sqrt{2}} \,
\left[\sum_{p=u,c}V_{pb} V_{pq}^* \left (C_1(\mu) O_1^p(\mu) +
C_2(\mu) O_2^p(\mu) \right) -V_{tb} V_{tq}^*\sum_{i=3}^{10} C_{i}(\mu) \,O_i(\mu)
\right] \;,
\en
where $q=d,s$. Here the Fermi constant $G_{F}=1.166 39\times
10^{-5} GeV^{-2}$, and the functions $Q_i (i=1,...,10)$ are the local four-quark operators. We specify below
the operators in ${\cal H}_{eff}$ for $b \to q$ transition: \be
\begin{array}{llllll}
O_1^{u} & = & \bar q_\alpha\gamma^\mu L u_\beta\cdot \bar
u_\beta\gamma_\mu L b_\alpha\ , &O_2^{u} & = &\bar
q_\alpha\gamma^\mu L u_\alpha\cdot \bar
u_\beta\gamma_\mu L b_\beta\ , \\
O_3 & = & \bar q_\alpha\gamma^\mu L b_\alpha\cdot \sum_{q'}\bar
 q_\beta'\gamma_\mu L q_\beta'\ ,   &
O_4 & = & \bar q_\alpha\gamma^\mu L b_\beta\cdot \sum_{q'}\bar
q_\beta'\gamma_\mu L q_\alpha'\ , \\
O_5 & = & \bar q_\alpha\gamma^\mu L b_\alpha\cdot \sum_{q'}\bar
q_\beta'\gamma_\mu R q_\beta'\ ,   & O_6 & = & \bar
q_\alpha\gamma^\mu L b_\beta\cdot \sum_{q'}\bar
q_\beta'\gamma_\mu R q_\alpha'\ , \\
O_7 & = & \frac{3}{2}\bar q_\alpha\gamma^\mu L b_\alpha\cdot
\sum_{q'}e_{q'}\bar q_\beta'\gamma_\mu R q_\beta'\ ,   & O_8 & = &
\frac{3}{2}\bar q_\alpha\gamma^\mu L b_\beta\cdot
\sum_{q'}e_{q'}\bar q_\beta'\gamma_\mu R q_\alpha'\ , \\
O_9 & = & \frac{3}{2}\bar q_\alpha\gamma^\mu L b_\alpha\cdot
\sum_{q'}e_{q'}\bar q_\beta'\gamma_\mu L q_\beta'\ ,   & O_{10} & =
& \frac{3}{2}\bar q_\alpha\gamma^\mu L b_\beta\cdot
\sum_{q'}e_{q'}\bar q_\beta'\gamma_\mu L q_\alpha'\ ,
\label{eq:operators} \end{array}
\en
where $\alpha$ and $\beta$ are
the $SU(3)$ color indices; $L$ and $R$ are the left- and
right-handed projection operators with $L=(1 - \gamma_5)$, $R= (1 +
\gamma_5)$. The sum over $q'$ runs over the quark fields that are
active at the scale $\mu=O(m_b)$, i.e., $(q'\epsilon\{u,d,s,c,b\})$.


\begin{figure}[t,b]
\vspace{-3cm} \centerline{\epsfxsize=16 cm \epsffile{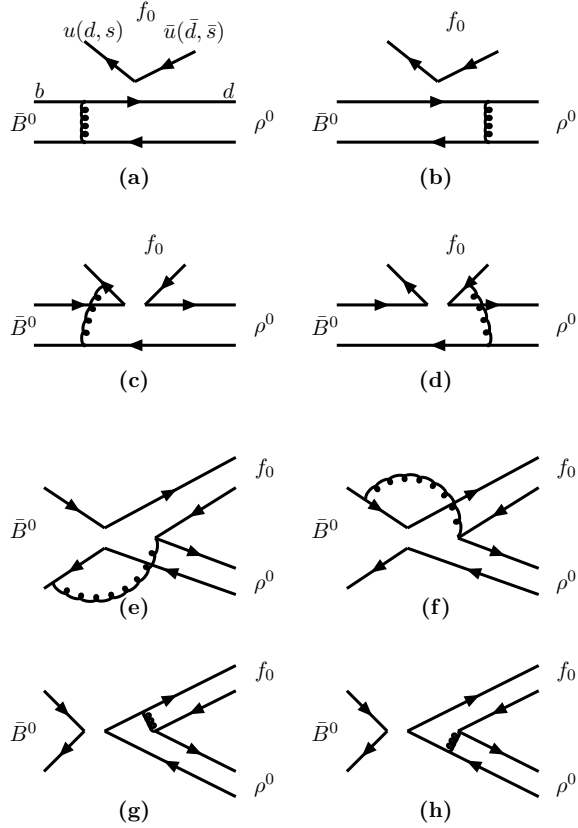}}
\vspace{-9cm} \caption{ Diagrams contributing to the decay $\bar{B}^0\to \rho^0
f_0(980)$ .}
 \label{fig1}
\end{figure}

In Fig.~1, we give the leading order Feynman diagrams for the channel $\bar{B}^0\to \rho^0 f_0(980)$ as an example. The
Feynman diagrams for the other decays are similar and not given in order to make a brief version. For the same purpose, the detailed analytic
formulae for the diagrams of each decays are not presented and can be gotten from those of $B\to f_0(980)K^*$ \cite{zqzhang} by
replacing corresponding wave functions and parameters.

Combining the contributions from different diagrams, the total decay
amplitudes for these decays can be written as:
\be {\cal M} (\bar
B\to f_0\rho(\phi,\omega)) &=&{\cal
M}_{s\bar{s}}(f_0\rho(\phi,\omega))\times\cos\theta+\frac{1}{\sqrt{2}}{\cal M}_{n\bar
n}(f_0\rho(\phi,\omega))\sin\theta, \en where $\theta$ is mixing angle and
\be
\sqrt{2}{\cal M}_{s\bar{s}}( f_0\rho^0)=-{\cal M}_{n\bar{n}}(f_0\rho^-)&=&\xi_t M_{e\rho}(C_4-\frac{1}{2}C_{10})+\xi_t M^{P2}_{e\rho}(C_6-\frac{1}{2}C_{8}),\label{eq:rho0}
\en
\be
\sqrt{2}{\cal M}_{s\bar{s}}( f_0\omega)&=&-\xi_t M_{e\rho}(C_4-\frac{1}{2}C_{10})-\xi_t M^{P2}_{e\rho}(C_6-\frac{1}{2}C_{8}),
\en
\be
{\cal M}_{s\bar{s}}(f_0\phi)&=&
-\xi_t \left[(M_{a\phi}+M_{af_0})(C_4-\frac{1}{2}C_{10})+(M^{P2}_{a\phi}+M_{af_0}^{af_0})
(C_6-\frac{1}{2}C_{8})\right.\non &&\left.+
(F_{af_0}+F_{a\phi})\left(a_3-a_5+\frac{1}{2}a_7-\frac{a_9}{2}\right)\right],\quad
\label{eq:omega}
\en
\be
\sqrt{2}{\cal M}_{n\bar{n}}(f_0\rho^0)&=&\bigg\{
\xi_u\left[ (M_{ef_0}+M_{af_0}-M_{e\rho}+M_{a\rho})C_2+(F_{ef_0}+F_{af_0}+F_{a\rho})a_2\right]\non &&
-\xi_t \left[F_{ef_0}(-a_4+\frac{3}{2}C_7+\frac{1}{2}C_8+\frac{5}{3}C_9+C_{10})+(F_{a\rho}+F_{af_0})\left(-a_4\right.\right.\non &&\left.\left.
-\frac{3}{2}C_7-\frac{1}{2}C_8+\frac{5}{3}C_9+C_{10}\right)-(F^{P2}_{a\rho}+F^{P2}_{af_0}+F^{P2}_{e\rho})(a_6-\frac{1}{2}a_8)\right.\non &&\left.
+(M_{ef_0}+M_{af_0}+M_{a\rho})(\frac{1}{2}C_9+\frac{3}{2}C_{10}-C_3)-M_{e\rho}\left(C_3+2C_4\right.\right.\non &&\left.\left.
-\frac{1}{2}C_9
+\frac{1}{2}C_{10}\right)-(M^{P1}_{ef_0}+M^{P1}_{e\rho}+M^{P1}_{af_0}+M^{P1}_{a\rho})(C_5-\frac{1}{2}C_7)\right.\non &&\left.-M^{P2}_{e\rho}(2C_6+\frac{1}{2}C_8)
+(M^{P2}_{ef_0}+M^{P2}_{af_0}+M^{P2}_{a\rho})
\frac{3}{2}C_8\right]\bigg\},\quad\;
\label{eq:rho0}
\en
\be
{\cal M}_{n\bar{n}}(f_0\rho^-)&=&\bigg\{
\xi_u\left[ M_{ef_0}C_2+(M_{af_0}+M_{e\rho}+M_{a\rho})C_1+F_{ef_0}a_2+(F_{af_0}+F_{a\rho})a_1\right]\non &&
-\xi_t \left[F^{P2}_{e\rho}(a_6-\frac{1}{2}a_8)+(F_{a\rho}+F_{ef_0}+F_{af_0})(a_4+a_{10})\right.\non &&\left.
+(F^{P2}_{a\rho}+F^{P2}_{af_0})(a_6+a_8)+(M_{ef_0}+M_{af_0}+M_{a\rho})(C_3+C_9)+M_{e\rho}\right.\non &&\left.
\times\left(C_3+2C_4-\frac{1}{2}C_9
+\frac{1}{2}C_{10}\right)+(M^{P1}_{ef_0}+M^{P1}_{af_0}+M^{P1}_{a\rho})\right.\non &&\left.\times(C_5+C_7)+M^{P1}_{e\rho}(C_5-\frac{1}{2}C_7)
+M^{P2}_{e\rho}(2C_6+\frac{1}{2}C_8)
\right]\bigg\},\quad\;
\label{eq:rhoz}
\en
\be
\sqrt{2}{\cal M}_{n\bar{n}}(f_0\omega)&=&\bigg\{
\xi_u\left[ (M_{ef_0}+M_{af_0}+M_{e\rho}+M_{a\rho})C_2+(F_{ef_0}+F_{af_0}+F_{a\rho})a_2\right]\non &&
-\xi_t \left[(F_{ef_0}+F_{a\rho}+F_{af_0})\left(\frac{7}{3}C_3+\frac{5}{3}C_4-2C_5-\frac{2}{3}C_6-\frac{1}{2}C_7\right.\right.\non &&\left.\left.
-\frac{1}{6}C_8+\frac{1}{3}C_9
-\frac{1}{3}C_{10}\right)+(F^{P2}_{a\rho}+F^{P2}_{af_0}+F^{P2}_{e\rho})(a_6-\frac{1}{2}a_8)\right.\non &&\left.
+(M_{e\rho}+M_{ef_0}+M_{af_0}+M_{a\rho})\left(C_3+2C_4-\frac{1}{2}C_9
+\frac{1}{2}C_{10}\right)\right.\non &&\left.
+(M^{P1}_{ef_0}+M^{P1}_{e\rho}+M^{P1}_{af_0}+M^{P1}_{a\rho})(C_5-\frac{1}{2}C_7)\right.\non &&\left.
+(M^{P2}_{ef_0}+M^{P2}_{af_0}+M^{P2}_{a\rho}+M^{P2}_{e\rho})(2C_6+\frac{1}{2}C_8)\right]\bigg\},\quad\;
\label{eq:omega}
\en
\be
{\cal M}_{n\bar{n}}(f_0\phi)&=&
-\xi_t \left[F_{ef_0}\left(a_3+a_5-\frac{a_7}{2}-\frac{a_9}{2}\right)+M_{ef_0}(C_4-\frac{C_{10}}{2})
+M^{P2}_{ef_0}(C_6-\frac{C_8}{2})\right];\quad
\label{eq:omega}
\en
\be
\sqrt{2}{\cal M}(K^{*-}\rho^0)&=&\xi_u\left[(M_{e\rho}+M_{a\rho})C_1+M_{eK^*}C_2+F_{a\rho}a_1
+F_{eK^*}a_2\right]-\xi_t\left[F^{P2}_{e\rho}(a_6+a_8)\right.\non &&\left.+F_{eK^*}(\frac{3}{2}(C_7+C_9)
+\frac{1}{2}(C_8+C_{10}))+(M_{e\rho}+M_{a\rho})(C_3+C_9)\right.\non &&\left.+(M^{P1}_{e\rho}+M^{P1}_{a\rho})
(C_5+C_7)+F_{a\rho}(a_4+a_{10})+F^{P2}_{a\rho}(a_6+a_{8})\right.\non &&\left.+M_{eK^*}\frac{3}{2}C_{10}
+M^{P2}_{eK^*}\frac{3}{2}C_{8}\right],
\en
\be
\sqrt{2}{\cal M}(K^{*-}\omega)&=&\xi_u\left[(M_{e\omega}+M_{a\omega})C_1+M_{eK^*}C_2+F_{a\omega}a_1
+F_{eK^*}a_2\right]-\xi_t\left[F^{P2}_{e\omega}(a_6+a_8)\right.\non &&\left.+F_{eK^*}(2(C_3+C_5)
+\frac{2}{3}(C_4+C_6)+\frac{1}{2}(C_9+C_7)
+\frac{1}{6}(C_8+C_{10}))\right.\non &&\left.+(M_{e\omega}+M_{a\omega})(C_3+C_9)+(M^{P1}_{e\omega}+M^{P1}_{a\omega})
(C_5+C_7)+F_{a\omega}(a_4+a_{10})\right.\non &&\left.+F^{P2}_{a\omega}(a_6+a_{8})+M_{eK^*}
(2C_4+\frac{1}{2}C_{10})+M^{P2}_{eK^*}(2C_6+\frac{1}{2}C_{8})\right],
\en
\be
{\cal M}(K^{*0}\rho^-)&=&\xi_u\left[M_{a\rho}C_1+F_{a\rho}a_1\right]-\xi_t\left[F^{P2}_{e\rho}(a_6+a_8)
+M_{e\rho}(C_3-\frac{C_9}{2})+M_{a\rho}(C_3+C_9)\right.\non &&\left.+M^{P1}_{e\rho}(C_5-\frac{C_7}{2})+M^{P1}_{a\rho}(C_5+C_7)
+F_{a\rho}(a_4+a_{10})+F^{P2}_{a\rho}(a_6+a_{8})\right],
\en
\be
\sqrt{2}{\cal M}(K^{*0}\rho^0)&=&\xi_u\left[M_{eK^*}C_2
+F_{eK^*}a_2\right]-\xi_t\left[F^{P2}_{e\rho}(a_6-\frac{1}{2}a_8)\right.\non &&\left.+F_{eK^*}(\frac{3}{2}(C_7+C_9)
+\frac{1}{2}(C_8+C_{10}))+(M_{e\rho}+M_{a\rho})(C_3-\frac{1}{2}C_9)\right.\non &&\left.+(M^{P1}_{e\rho}+M^{P1}_{a\rho})
(C_5-\frac{1}{2}C_7)+F_{a\rho}(a_4-\frac{1}{2}a_{10})+F^{P2}_{a\rho}(a_6-\frac{1}{2}a_{8})\right.\non &&\left.+M_{eK^*}\frac{3}{2}C_{10}
+M^{P2}_{eK^*}\frac{3}{2}C_{8}\right],
\en
\be
\sqrt{2}{\cal M}(K^{*0}\omega)&=&\xi_u\left[M_{eK^*}C_2
+F_{eK^*}a_2\right]-\xi_t\left[F^{P2}_{e\omega}(a_6-\frac{1}{2}a_8)+(M_{e\omega}+M_{a\omega})(C_3-\frac{1}{2}C_9)\right.\non &&\left.
+F_{eK^*}(2(C_3+C_5)
+\frac{2}{3}(C_4+C_6)+\frac{1}{2}(C_9+C_7)
+\frac{1}{6}(C_8+C_{10}))\right.\non &&\left.+(M^{P1}_{e\omega}+M^{P1}_{a\omega})
(C_5-\frac{1}{2}C_7)+F_{a\omega}(a_4-\frac{1}{2}a_{10})+F^{P2}_{a\omega}(a_6-\frac{1}{2}a_{8})\right.\non &&\left.+M_{eK^*}(2C_4+\frac{1}{2}C_{10})
+M^{P2}_{eK^*}(2C_6+\frac{3}{2}C_{8})\right],
\en
\be
{\cal M}(K^{*-}\rho^+)&=&\xi_uM_{e\rho}C_1-\xi_t\left[F^{P2}_{e\rho}(a_6+a_8)
+M_{e\rho}(C_3+C_9)+M_{a\rho}(C_3-\frac{C_9}{2})\right.\non &&\left.+M^{P1}_{e\rho}(C_5+C_7)+M^{P1}_{a\rho}(C_5-\frac{C_7}{2})
+F_{a\rho}(a_4-\frac{a_{10}}{2})+F^{P2}_{a\rho}(a_6-\frac{a_{8}}{2})\right].\qquad
\en
The
combinations of the Wilson coefficients are defined as usual
\cite{zjxiao}:
 \be
a_{1}(\mu)&=&C_2(\mu)+\frac{C_1(\mu)}{3}, \quad
a_2(\mu)=C_1(\mu)+\frac{C_2(\mu)}{3},\non
a_i(\mu)&=&C_i(\mu)+\frac{C_{i+1}(\mu)}{3},\quad
i=3,5,7,9,\non
a_i(\mu)&=&C_i(\mu)+\frac{C_{i-1}(\mu)}{3},\quad
i=4, 6, 8, 10.\label{eq:aai} \en

\section{Numerical results and discussions} \label{numer}

We use the following input parameters in the numerical calculations \cite{pdg08,barbar}:
\be
f_B&=&190 MeV, M_B=5.28 GeV, M_W=80.41 GeV,\\
V_{ub}&=&|V_{ub}|e^{-i\gamma}=3.93\times10^{-3}e^{-i68^\circ}, V_{tb}=1.0,\\
 V_{td}&=&|V_{ud}|e^{-i\beta}=8.1\times10^{-3}e^{-i21.6^\circ},V_{us}=0.2255,\\
 V_{ts}&=&0.0387,V_{ud}=0.974, \tau_{B^\pm}=1.671\times 10^{-12} s,\\
\tau_{B^0}&=&1.530\times 10^{-12} s.
\en .

In the B-rest frame, the decay rates of $B\to f_0(980)\rho(\omega,\phi), K^*_0(1430)\rho(\omega)$ can be written as:
\be
\Gamma=\frac{G_F^2}{32\pi m_B}|{\cal M}|^2(1-r^2_S),
\en
where ${\cal M}$ is the total decay amplitude of each
considered decay and $r_S$ the mass ratio, which have been given  in  section \ref{results}.

If $f_0(980)$ is purely composed of $n\bar n$($s\bar s$), the branching ratios
of $B^-\to f_0(980)\rho^-$ and $\bar B^0\to f_0(980)\rho^0(\omega,\phi)$ are:
\be
{\cal B}(B^-\to
f_0(980)(n\bar n)\rho^-)&=&(7.5^{+0.9+1.4+1.4}_{-0.8-1.1-1.1})\times 10^{-6},\\
{\cal B}(\bar B^0\to
f_0(980)(n\bar n)\rho^0)&=&(1.1^{+0.2+0.3+0.2}_{-0.1-0.2-0.3})\times 10^{-6},\\
{\cal B}(\bar B^0\to
f_0(980)(n\bar n)\omega)&=&(5.3^{+0.5+1.1+0.9}_{-0.5-0.9-0.6})\times 10^{-6},\\
{\cal B}(\bar B^0\to
f_0(980)(n\bar n)\phi)&=&(1.7^{+0.2+0.5+0.3}_{-0.2-0.4-0.3})\times 10^{-9},\\
{\cal B}(B^-\to
f_0(980)(s\bar s)\rho^-)&=&(3.0^{+0.3+0.7+0.5}_{-0.3-0.6-0.4})\times 10^{-7},\\
{\cal B}(\bar B^0\to
f_0(980)(s\bar s)\rho^0)&=&(1.4^{+0.3+0.3+0.2}_{-0.2-0.3-0.2})\times 10^{-7},\\
{\cal B}(\bar B^0\to
f_0(980)(s\bar s)\omega)&=&(1.2^{+0.1+0.3+0.2}_{-0.1-0.2-0.2})\times 10^{-7},\\
{\cal B}(\bar B^0\to
f_0(980)(s\bar s)\phi)&=&(2.0^{+0.2+0.4+0.0}_{-0.2-0.3-0.1})\times 10^{-8},
\end{eqnarray}
where the uncertainties are from the decay constant of $f_0(980)$,
the Gegenbauer moments $B_1$,$B_3$ and the $B$ meson shape parameter $\omega=0.40\pm0.04$ GeV. In these
$b\to d$ transition processes, the decay $\bar B^0\to
f_0(980)\phi$ is very different from the other three channels: the value of ${\cal B}(B\to f_0(n\bar n)\phi)$
is smaller than that of ${\cal B}(\bar B\to f_0(s\bar s)\phi)$
about one order, it is contrary to the cases of the other three decays, at the same time, the branching ratios for $n\bar n$
and $s\bar s$ components of this channel are both very small.

\begin{figure}[t,b]
\begin{center}
\includegraphics[scale=0.7]{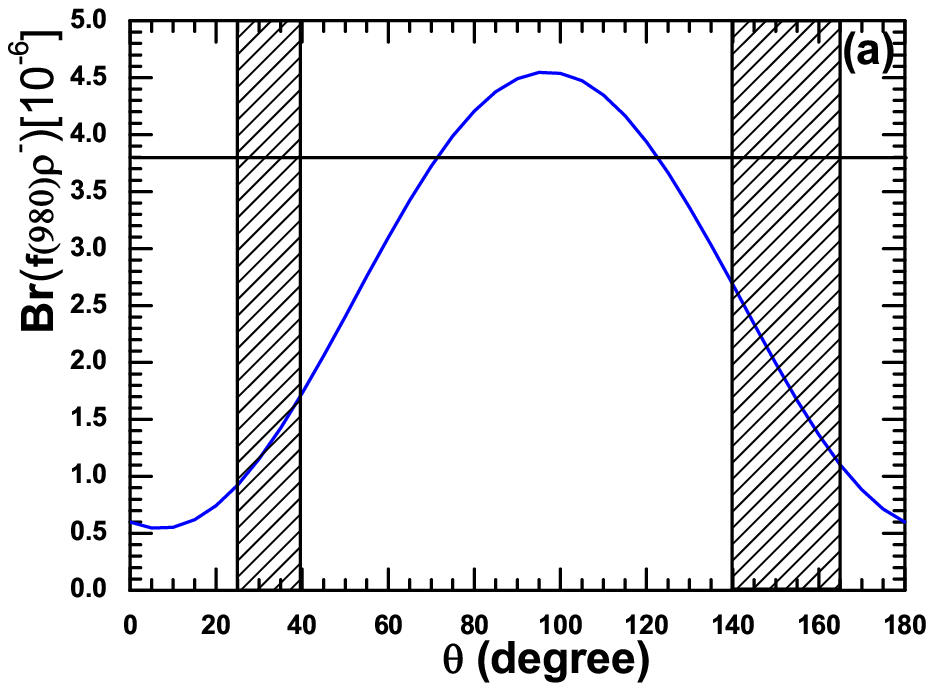}
\includegraphics[scale=0.7]{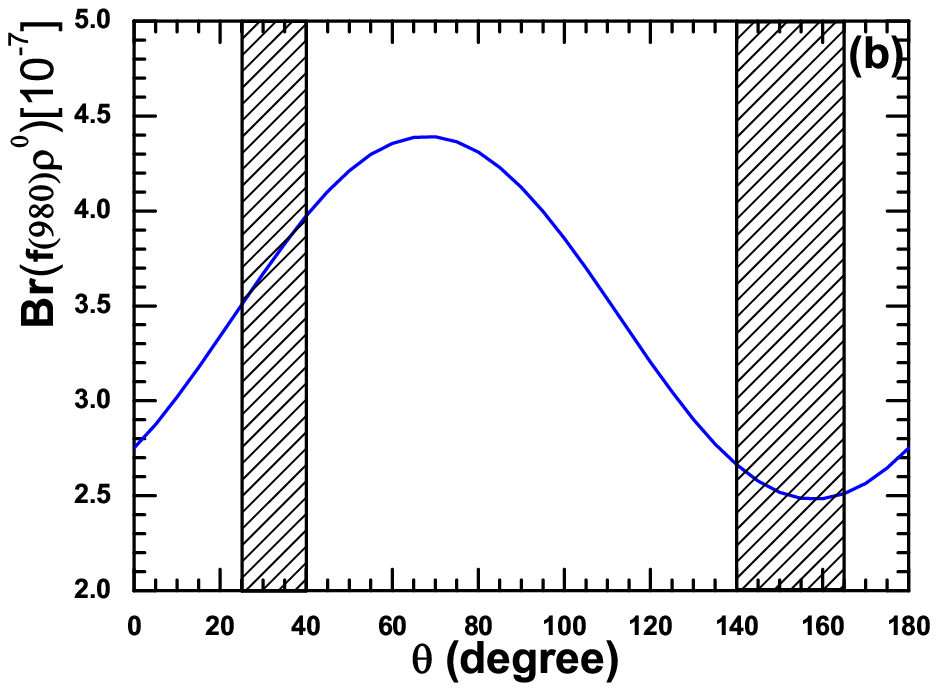}
\includegraphics[scale=0.7]{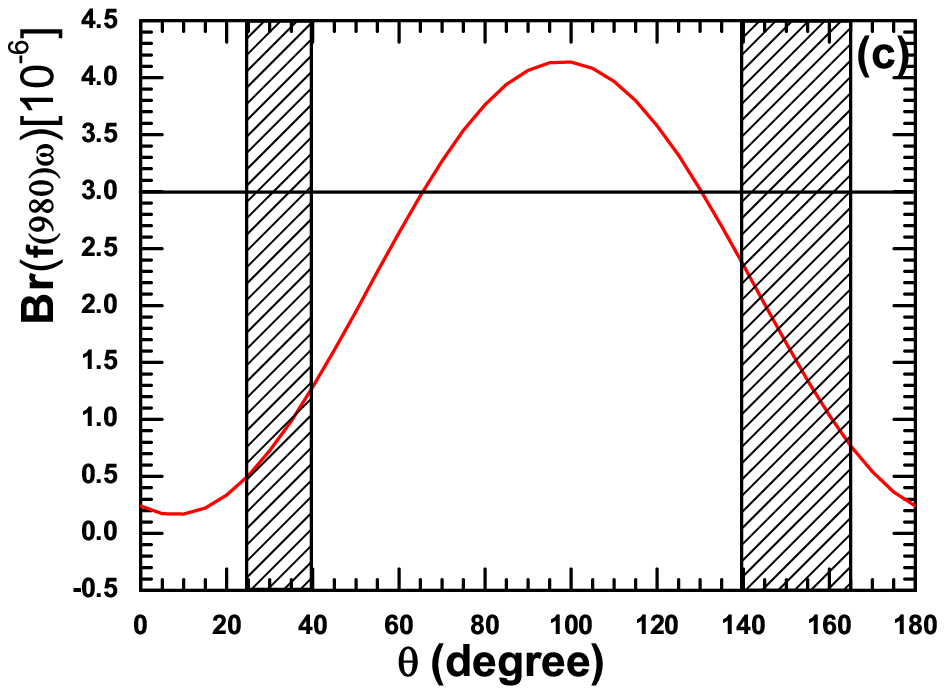}
\includegraphics[scale=0.7]{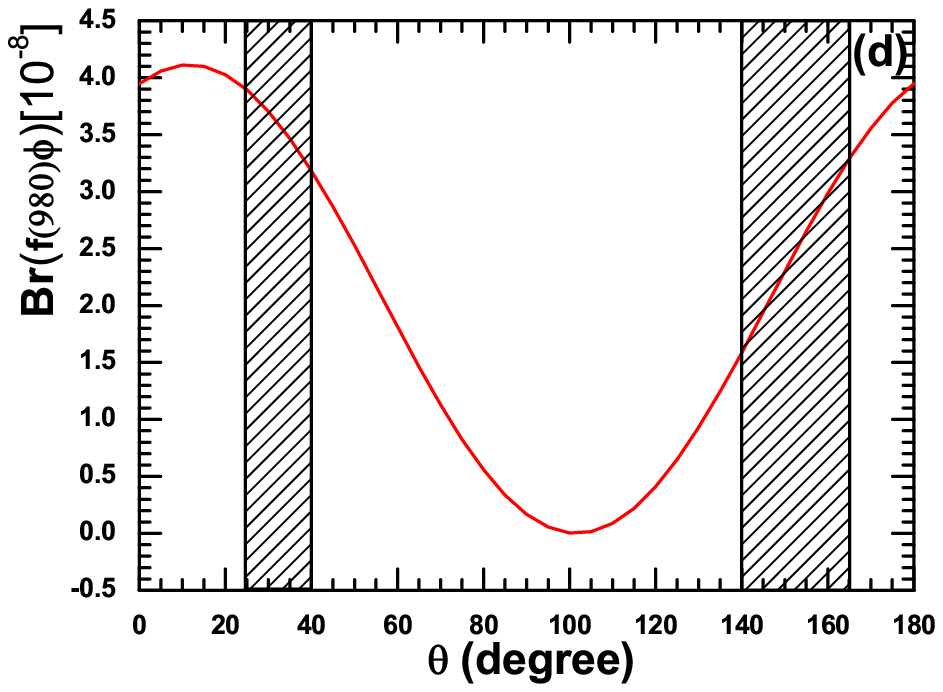}
\vspace{0.3cm} \caption{The dependence of the branching ratios for
$B^-_s\to f_0(980)\rho^-$ (a), $\bar B^0\to f_0(980)\rho^0$ (b), $\bar B^0\to f_0(980)\omega$ (c) and
$\bar B^0\to f_0(980)\phi$ (a) on the
mixing angle $\theta$ using the inputs derived from QCD sum rules. The horizontal solid lines show
the experimental upper limits.
The vertical bands show
two possible ranges of $\theta$: $25^\circ<\theta<40^\circ$ and $140^\circ<\theta<165^\circ$.}\label{fig2}
\end{center}
\end{figure}
In Fig.~\ref{fig2}, we plot the branching ratios of the considered decays as functions of
the mixing angle $\theta$. One can find our predictions for the decays $B^-\to f_0(980)\rho^-$
and $\bar B^0\to f_0(980)\omega$ are smaller than the experimental upper limits, but not far away from them. In these decay channels,
the branch ratio of $B^-\to f_0(980)\rho^-$ is the largest
one, most possible in the range $(1.0\sim2.5)\times10^{-6}$. We predict that the branch
ratio of the decay $\bar B^0\to f_0(980)\rho^0$ is at the order of $10^{-7}$. The tree operator contributions of different
diagrams are destructive inference, which leads the tree dominated decay $\bar B^0\to f_0(980)\rho^0$ to receive a rather small rate.
On the contrary, the different amplitudes of the decay $\bar B^0\to f_0(980)\omega$ are constructive inference and this channel has a larger rate,
which is close to the branch ratio of
$B^-\to f_0(980)\rho^-$. Certainly, this scheme (the inference between difference tree contributions) is influenced by the value
of the mixing angle, for example, it is not obvious for
$\theta=20^\circ$, while obvious for $25^\circ<\theta<40^\circ$ and $140^\circ<\theta<165^\circ$.
As to the decay $\bar B^0\to f_0(980)\phi$, there are no tree contributions
in the leading order and the contributions from the $s\bar s$ component are document. One can see that its branching ratio
is very small and has a different dependence on the mixing angle with other three decays.
Its theoretical value is in the range
\be
2.2\times 10^{-8}&<&{\cal B}(\bar B^0\to f_0(980)\rho)<3.8\times 10^{-8}, \quad\mbox{for} \quad 25^\circ<\theta<40^\circ;\\
4.6\times 10^{-8}&<&{\cal B}(\bar B^0\to f_0(980)\rho)<6.0\times 10^{-8},\quad \mbox{for}\quad 140^\circ<\theta<165^\circ,
\en
which is far smaller than its upper limit $38\times10^{-8}$.

For comparison, we also give the theoretical
results in the QCDF framework \cite{ccysv}, which are listed in Table I. Obviously, there exists stark disagreement with the QCDF
predictions. It mainly arises from taking different values about the decay constants of the scalar mesons and dealing with the
annihilation diagram contributions in different way.

\begin{table}
\caption{Branching ratios (in units of $10^{-6}$) of $B\to f_0(980)\rho(\omega, \phi)$ and $B\to K^*_0(1430)\rho(\omega)$. The theoretical
errors correspond to the uncertainties due to (i) the scalar meson decay constants, (ii) the Gegenbauer moments $B_1$ and $B_3$ for the
scalar mesons, (iii) the $B$ meson shape parameter $\omega$. In order to compare with the QCDF predictions, we also give the predicted branching ratios
of $B\to f_0(980)\rho(\omega, \phi)$ for the $f_0-\sigma$ mixing angle $\theta=20^\circ$. For the QCDF results, the branching ratios
of $B\to f_0(980)V$ are in SI, ones of $B\to K^*_0(1430)V$ are in SII.}\label{para}
\begin{center}
\begin{tabular}{c|c|c|c|c}
\hline \hline
 Mode& QCDF& scenario I& scenario II&Exp.\\
 \hline
$B^-\to f_0(980)\rho^-$& $1.3^{+0.1+0.4+0.1}_{-0.1-0.3-0.1}$ &$0.7^{+0.1+0.2+0.2}_{-0.0-0.1-0.1}$ & & $<3.8$ \\
$\bar B^0\to f_0(980)\rho^0$& $0.01^{+0.00+0.00+0.02}_{-0.00-0.00-0.01}$&$0.33^{+0.04+0.07+0.06}_{-0.03-0.05-0.06}$& & $<1.06$ \\
$\bar B^0\to f_0(980)\phi$ &--&$0.04^{+0.005+0.008+0.000}_{-0.004-0.007-0.003}$& &$<0.76$\\
 $\bar B^0\to f_0(980)\omega$ &$0.06^{+0.02+0.00+0.02}_{-0.01-0.00-0.02}$&$0.34^{+0.03+0.06+0.06}_{-0.04-0.06-0.05}$& &$<3.0$\\
 \hline
$B^-\to \bar K^{*0}_0(1430)\rho^-$& $66.2^{+25.0+2.8+70.8}_{-19.5-2.4-26.3}$&$3.2^{+0.7+0.8+0.4}_{-0.6-0.7-0.2}$& $12.1^{+2.8+3.9+0.5}_{-0.0-3.1-0.5}$&\\
$B^-\to K^{*-}_0(1430)\rho^0$ &$21.0^{+7.3+1.2+29.4}_{-5.9-1.1-10.1}$ &$3.4^{+0.8+0.7+0.6}_{-0.6-0.5-0.4}$& $8.4^{+2.3+3.3+0.9}_{-0.0-3.2-0.7}$&\\
$B^-\to K^{*-}_0(1430)\omega$ &$16.1^{+4.9+0.7+22.5}_{-4.0-0.6-8.3}$&$3.8^{+0.9+0.5+0.8}_{-0.9-0.6-0.7}$& $7.4^{+2.1+3.0+0.9}_{-1.5-2.3-0.4}$&\\
$\bar B^0\to \bar K^{*0}_0(1430)\rho^0$& $36.8^{+14.3+0.9+23.4}_{-11.0-0.7-9.1}$&$0.47^{+0.12+0.20+0.03}_{-0.12-0.17-0.02}$& $4.8^{+1.1+1.0+0.3}_{-0.0-1.0-0.3}$&\\
$\bar B^0\to  K^{*-}_0(1430)\rho^+$& $51.0^{+16.1+1.4+68.6}_{-13.1-1.2-23.8}$&$3.3^{+0.7+0.8+0.2}_{-0.6-0.8-0.2}$&$10.5^{+2.7+3.5+0.3}_{-0.0-2.6-0.3} $&\\
$\bar B^0\to K^{*0}_0(1430)\omega$ &$15.6^{+4.4+1.0+14.6}_{-3.7-0.8-5.3}$&$4.9^{+1.2+0.7+1.1}_{-1.1-0.7-0.9}$& $9.3^{+2.1+3.6+1.2}_{-2.0-2.9-1.0}$&\\
\hline \hline
\end{tabular}
\end{center}
\end{table}
\begin{figure}[t,b]
\begin{center}
\includegraphics[scale=0.7]{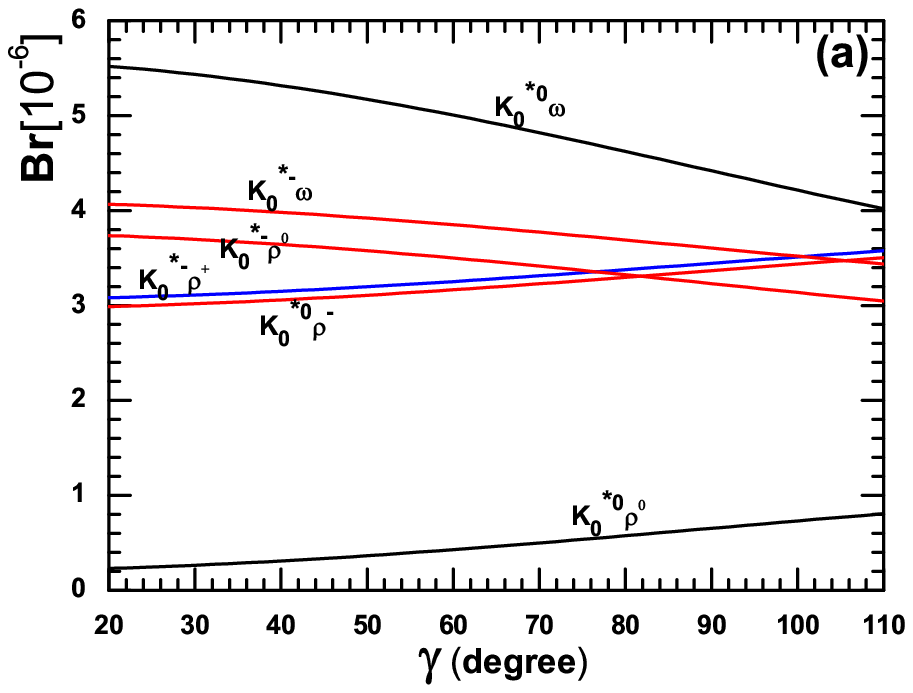}
\includegraphics[scale=0.7]{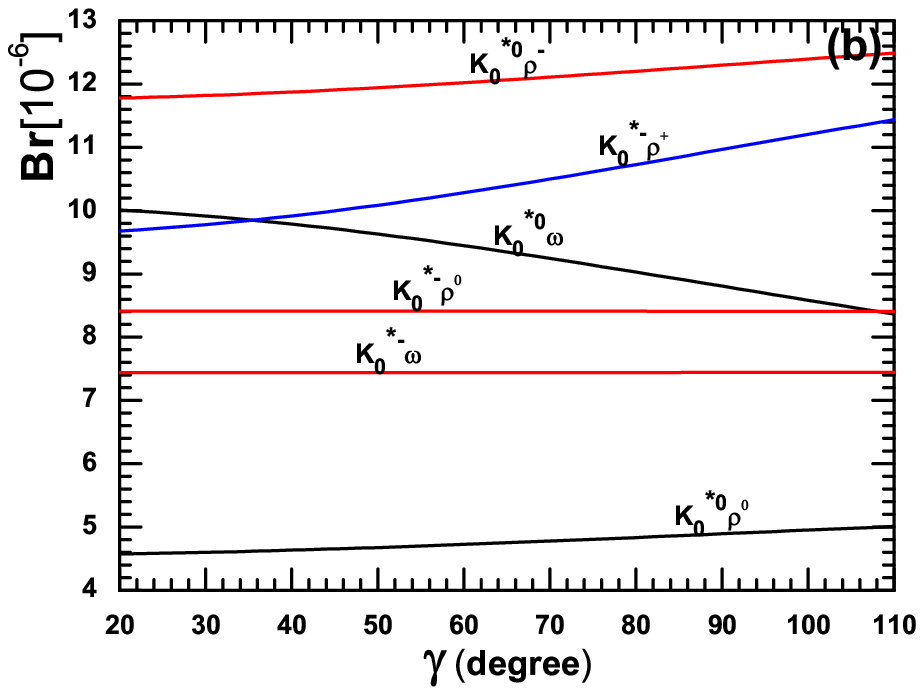}
\vspace{0.3cm} \caption{The dependence of the branching ratios for
$B^-\to K^{*0}_0\rho^-, K^{*-}_0\rho^0, K^{*-}_0\omega$ and  $\bar B^0\to K^{*-}_0\rho^+, K^{*0}_0\rho^0, K^{*0}_0\omega$ on the
CKM angle $\gamma$.}\label{fig3}
\end{center}
\end{figure}
\begin{table}
\caption{ Decay amplitudes for decays $B^-\to K^{*-}_0(1430)\omega, \bar B^0\to K^{*0}_0(1430)\omega$ ($\times 10^{-2} \mbox {GeV}^3$).}
\begin{center}
\begin{tabular}{cc|c|c|c|c|c|c|c}
\hline \hline  &&$F_{e\omega}$ & $M^T_{e\omega}$ &$M_{e\omega}$& $M^T_{a\omega}$ & $M_{a\omega}$ &$F^T_{a\omega}$\\
\hline
$B^-\to K^{*-}_0(1430)\omega$ (SI) &   &0.6&$13.2-18.4i$&$-0.5+0.7i$&$12.3-12.0i$&$-0.3+0.5i$ &$-1.7-2.2i$\\
$\bar B^0\to K^{*0}_0(1430)\omega$ (SI) &   &0.6&--&$-0.8+1.2i$&--&$-0.7+0.8i$ &--\\
$B^-\to K^{*-}_0(1430)\omega$ (SII) &   &-0.9&$-17.3-14.6i$&$0.6+0.5i$&$-13.8-0.8i$&$0.5-0.1i$ &$-0.2+0.3i$\\
$\bar B^0\to K^{*0}_0(1430)\omega$ (SII) &   &-0.9&--&$1.1+0.9i$&--&$0.9-0.1i$ &--\\
\hline   &&$F_{a\omega}$ & $F^T_{eK^*}$ &$F_{eK^*}$& $M^T_{eK^*}$ & $M_{eK^*}$ &$F^T+M^T$\\
\hline
$B^-\to K^{*-}_0(1430)\omega$ (SI) &   &$-3.5-2.2i$&-32.4&-2.3&$-11.0+6.3i$&$-0.6+0.3i$ &$-18.5-25.6i$\\
$\bar  B^0\to K^{*0}_0(1430)\omega$ (SI) &   &$-3.5-2.0i$&-32.4&-2.3&$-11.0+6.3i$&$-0.6+0.3i$ &$-43.4+6.3i$\\
$B^-\to K^{*-}_0(1430)\omega$ (SII) &   &$2.9+5.9i$&41.2&3.1&$0.1+6.1i$&$0.0+0.4i$ &$10.0-9.0i$\\
$\bar  B^0\to K^{*0}_0(1430)\omega$ (SII) &   &$3.1+6.2i$&41.2&3.1&$0.1+6.1i$&$0.0+0.4i$ &$41.3+6.1i$\\
\hline \hline
\end{tabular}\label{amp}
\end{center}
\end{table}
As to the decays $B^-\to K^{*0}_0\rho^-, K^{*-}_0\rho^0, K^{*-}_0\omega$ and $\bar B^0\to K^{*-}_0\rho^+, K^{*0}_0\rho^0,
K^{*0}_0\omega$,
though there are no the experimental results, our argument is that
the branch ratios of decays $B\to K^*_0(1430)\rho(\omega)$ might not far away from those of
$B\to K\rho(\omega)$, just like the relationship between $B\to K^*_0(1430)\phi$ and $B\to K\phi$ \cite{yingli}. It is not like the channels
$B\to f_0(980)\rho(\omega)$, where there exists large destructive (constructive) conference between the components $u\bar u$ and $d\bar d$ in
the mesons $f_0(980)$ and $\rho^0$ ($\omega$), there exists relatively small conference in
decays $B\to K^*_0(1430)\rho(\omega)$, so the branching ratios of these decays are close to each other, most of them are in the range of
$(3\sim5)\times10^{-6}$ for scenario I, $(7\sim10)\times{10^{-6}}$ for scenario II. The branch ratio of $\bar B^0\to K^{*0}_0(1430)\rho^0$ is the smallest one in these decays, its value is at
the order of $10^{-7}$ in scenario I. Certainly, we only calculate the leading order diagrams, and do not consider the higher order corrections. If the
future experimental value about this channel is larger than our prediction, say $10^{-6}$, it indicates that this decay might be
much sensitive
to next leading order corrections, it is similar to the decays $B^0\to \pi^0\pi^0, \rho^0\rho^0$. On the contrary, the decay
$\bar B^0\to K^{*0}_0(1430)\omega$ arrives at a large rate in our leading order calculations, especially in scenario I. We expect that its value
will be
smaller after considering next leading order corrections. In Table II, we list the values of the factorizable and non-factorizable amplitudes
from the emission and annihilation topology diagrams of the decays $B^-\to K^{*-}_0(1430)\omega$ and $\bar B^0\to K^{*0}_0(1430)\omega$.
$F_{e(a)\omega}$ and $M_{e(a)\omega}$
are the $K^*_0(1430)$ emission (annihilation) factorizable
contributions and non-factorizable contributions from penguin operators respectively. Similarly,
$F_{e(a)K^*_0}$ and $M_{e(a)K^*_0}$ denote
the contributions from $\omega$ emission (annihilation) factorizable contributions and non-factorizable
contributions from penguin operators respectively. The upper label "T" denotes the
contributions from tree operators. For the $\omega$ emission type diagrams, these two decays have the same Wilson coefficients, so
the corresponding amplitudes have the same values.
The column "$F^T+M^T$" is for the
total tree contribution of  factorizable and non-factorizable diagrams.
From Table II, one can find that the tree contributions from $\omega$ and $K^*$ emission
type diagrams
are destructive in the charged decay, and a smaller real part of the total tree contribution survives in compare with the neutral one, which makes the
branching ratio of $\bar B^0\to K^{*0}_0(1430)\omega$ is larger than that of $B^-\to K^{*-}_0(1430)\omega$.

Now we turn to the evaluations of the direct CP-violating asymmetries of
the considered decays in PQCD approach. The direct CP-violating asymmetry can be defined as:
\be
\acp^{dir}=\frac{ |\overline{\cal M}|^2-|{\cal M}|^2 }{
 |{\cal M}|^2+|\overline{\cal M}|^2}\;.
\en

\begin{figure}[t,b]
\begin{center}
\includegraphics[scale=0.7]{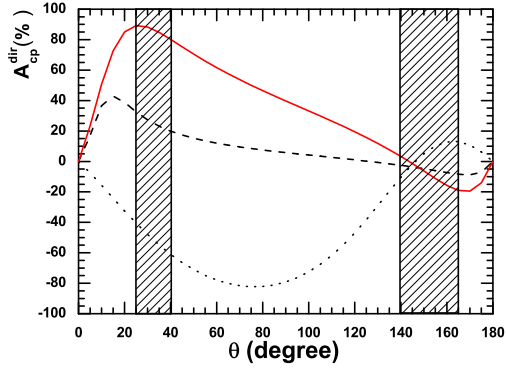}
\vspace{0.3cm} \caption{The dependence of the direct CP asymmetries for
$B^-\to f_0(980)\rho^-$ (solid curve), $\bar B^0\to f_0(980)\rho^0$ (dotted curve), $\bar B^0\to f_0(980)\omega$ (dashed curve) on the
mixing angle $\theta$.
The vertical bands show
two possible ranges of $\theta$: $25^\circ<\theta<40^\circ$ and $140^\circ<\theta<165^\circ$.}\label{fig2}
\end{center}
\end{figure}
\begin{figure}[t,b]
\begin{center}
\includegraphics[scale=0.7]{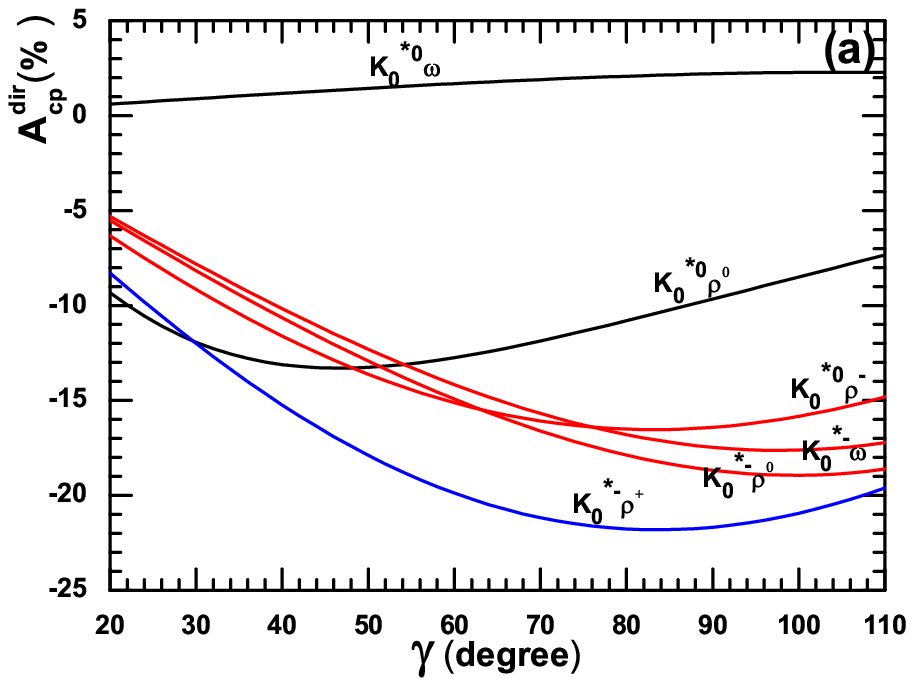}
\includegraphics[scale=0.7]{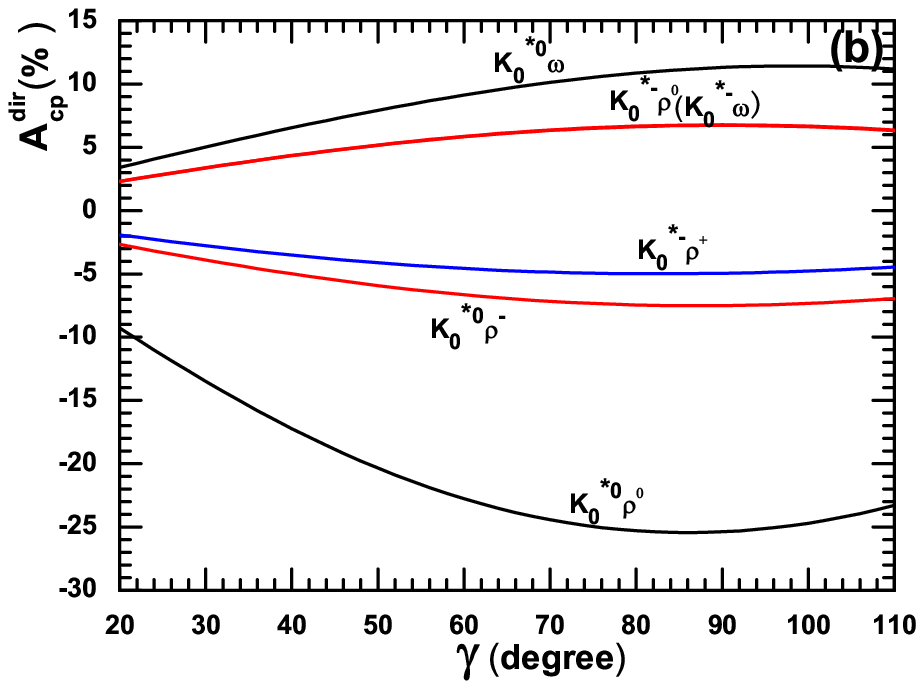}
\vspace{0.3cm} \caption{The dependence of the direct CP-violating asymmetries for
$B^-\to K^{*0}_0\rho^-, K^{*-}_0\rho^0, K^{*-}_0\omega$ and  $\bar B^0\to K^{*-}_0\rho^+, K^{*0}_0\rho^0, K^{*0}_0\omega$ on the
CKM angle $\gamma$.}\label{fig2}
\end{center}
\end{figure}
\begin{table}
\caption{Direct CP-violating asymmetries (in units of $\%$) of $B\to f_0(980)\rho(\omega, \phi)$ and $B\to K^*(1430)\rho(\omega)$. The errors
for these entries correspond to the uncertainties from the scalar meson decay constants, the Gegenbauer moments $B_1$ and $B_3$ for
the scalar meson and the $B$ meson shape parameter.
Here we still give the predicted direct CP asymmetries
of $B\to f_0(980)\rho(\omega, \phi)$ for the $f_0-\sigma$ mixing angle $\theta=20^\circ$.}\label{para}
\begin{center}
\begin{tabular}{c|c|c}
\hline \hline
 Mode& scenario I& scenario II\\
  \hline
$B^-\to f_0(980)\rho^-$&$85.1^{+0.0+1.8+4.7}_{-0.0-1.8-4.9}$ &  \\
$\bar B^0\to f_0(980)\rho^0$&$-32.4^{+0.0+8.2+3.3}_{-0.0-8.8-8.5}$& \\
$\bar B^0\to f_0(980)\phi$ &0& \\
 $\bar B^0\to f_0(980)\omega$ &$38.7^{+6.5+6.4+12.0}_{-0.0-7.7-10.3}$& \\
 \hline
$B^-\to \bar K^{*0}_0(1430)\rho^-$& $-15.9^{+0.0+1.0+0.8}_{-0.0-1.3-0.8}$&$-7.1^{+0.0+1.6+0.2}_{-0.0-1.0-0.2}$\\
$B^-\to K^{*-}_0(1430)\rho^0$ &$-16.3^{+0.0+1.8+2.7}_{-0.0-2.0-2.2}$ &$6.3^{+0.0+3.9+2.9}_{-0.1-3.8-2.6}$\\
$B^-\to K^{*-}_0(1430)\omega$ &$-15.4^{+0.2+1.5+2.6}_{-0.3-1.6-2.4}$&$6.2^{+0.0+4.2+3.0}_{-0.0-3.4-2.6}$\\
$\bar B^0\to \bar K^{*0}_0(1430)\rho^0$& $-12.1^{+0.0+8.5+2.0}_{-0.0-7.8-5.6}$&$-24.2^{+0.2+4.6+3.8}_{-0.0-2.7-4.0}$\\
$\bar B^0\to  K^{*-}_0(1430)\rho^+$& $-21.0^{+0.0+2.5+1.1}_{-0.0-2.6-0.7}$&$-4.8^{+0.3+0.9+0.4}_{-0.0-0.9-0.5}$\\
$\bar B^0\to K^{*0}_0(1430)\omega$ &$1.9^{+0.0+0.7+0.6}_{-0.0-0.7-0.0}$&$10.0^{+0.1+2.7+0.9}_{-0.0-2.5-0.9}$\\
\hline \hline
\end{tabular}
\end{center}
\end{table}
From Fig.4, one can see the direct CP-violating
asymmetry values for the decays $B^-\to f_0(980)\rho^-$ and $\bar B^0\to f_0(980)\rho^0$ in these two possible ranges of
the mixing angle $\theta$ are very different, that is to say, their CP-violating asymmetries are sensitive to the mixing angle. For the
decay $\bar B^0\to f_0(980)\omega$, its CP-violating asymmetry is not so sensitive to the mixing angle.
If the mixing angle is in the range $25^\circ<\theta<40^\circ$,
the direct CP-violating asymmetries of these decays are about
\be
80\%<{\cal A}^{dir}_{CP}(B^-\to f_0(980)\rho^-)<90\%,\\
-60\%<{\cal A}^{dir}_{CP}(\bar B^0\to f_0(980)\rho^0)<-40\%,\\
20\%<{\cal A}^{dir}_{CP}(\bar B^0\to f_0(980)\omega)<35\%.
\en
If the mixing angle is in the range $140^\circ<\theta<165^\circ$, the direct CP-violating asymmetries of these decays are about
\be
-20\%<{\cal A}^{dir}_{CP}(B^-\to f_0(980)\rho^-)<5\%,\\
-12\%<{\cal A}^{dir}_{CP}(\bar B^0\to f_0(980)\rho^0)<15\%,\\
-10\%<{\cal A}^{dir}_{CP}(\bar B^0\to f_0(980)\omega)<4\%.
\en

Certainly, we consider that the gluon component is small and neglectable in the meson $f_0(980)$. Our argument is that the
neglected gluon component has a small influence on the branching ratio, while has a bit more influence on the CP-violating asymmetry. So
if the contribution from gluon content is included, it will give these direct CP-violating asymmetry values some
corrections. As to the decay $\bar B^0\to f_0(980)\phi$, there is no tree contribution at the
leading order, so the direct CP-violating asymmetry is naturally zero.

For the decay $\bar B^0\to K^{*0}_0\omega$, its direct CP-violating asymmetries for two scenarios are both positive and small.
For the charged decays $B^-\to \bar K^{*0}_0(1430)\rho^-, K^{*-}_0(1430)\rho^0, K^{*-}_0(1430)\omega$, their direct CP asymmetries have similar size in two scenarios. While in scenario II, the
decay $B^-\to \bar K^{*0}_0(1430)\rho^-$, which branching ratio is the biggest one, has an opposite sign
with the other two charged decays. It is because that there exist contributions from the vector meson emission
diagrams in the decays $B^-\to K^{*-}_0(1430)\rho^0, K^{*-}_0(1430)\omega$, which will flip the signs of their direct
CP-violating asymmetry values when the wave function of $K^{*}_0(1430)$ in scenario II is used, while there are not these kinds of extra contributions
in the decay $B^-\to \bar K^{*0}_0(1430)\rho^-$.


\section{Conclusion}\label{summary}

In this paper, we calculate the branching ratios and the CP-violating
asymmetries of decays $B\to f_0(980)\rho(\omega,\phi), K^*_0(1430)\rho(\omega)$
in the PQCD factorization approach.
Using the decay constants and light-cone distribution amplitudes
derived from QCD sum-rule method, we find that:
\begin{itemize}
\item
If $f_0(980)$ is purely composed of $n\bar n$($s\bar s$), the value of ${\cal B}(B\to f_0(n\bar n)\phi)$
is smaller than that of ${\cal B}(\bar B\to f_0(s\bar s)\phi)$
about one order for the channel $\bar B\to f_0(s\bar s)\phi$ (it is contrary to the cases of $\bar B^0\to
f_0(980)\rho(\omega)$), at the same time, these two branching ratios for $n\bar n$
and $s\bar s$ components are both very small.

\item
In the $b\to d$ transition processes $B\to f_0(980)\rho(\omega,\phi)$, the branch ratio
of $B^-\to f_0(980)\rho^-$
is the largest one and its value is possible in the range $(1.0\sim2.5)\times10^{-6}$, the branch
ratio of $\bar B^0\to f_0(980)\rho^0$ is at the order of $10^{-7}$. Our predictions for the decays $B^-\to f_0(980)\rho^-$
and $\bar B^0\to f_0(980)\omega$ are smaller than the experimental upper limits, but not far away from them.

\item
In the $b\to s$ transition processes $B\to K^{*}_0\rho(\omega)$, there exists small difference for the values of their branch ratios, most of them are in the range of
$(3\sim5)\times10^{-6}$ for scenario I, $(7\sim10)\times{10^{-6}}$ for scenario II.

\item
In scenario I, the branch ratio of $\bar B^0\to K^{*0}_0(1430)\rho^0$ is the smallest one in these $b\to s$ transition processes,
its value is at the order of $10^{-7}$ in scenario I. Certainly, we only calculate the leading order diagrams, and do not consider the higher
order corrections. If the future experimental value about this channel is larger than our prediction, say $10^{-6}$,
it indicates that this decay might be more sensitive
to next leading order corrections, which is similar to the decays $B^0\to \pi^0\pi^0, \rho^0\rho^0$. On the other side, the decay
$\bar B^0\to K^{*0}_0(1430)\omega$ arrives at a large rate in our leading order calculations. We expect that its value
will be smaller after considering next leading order corrections.
\item
The direct CP-violating asymmetries of decays $B\to f_0(980)\rho(\omega)$ have a strong dependent on the mixing angle $\theta$: they are large
in the range of $25^\circ<\theta<40^\circ$, and small in the range of $140^\circ<\theta<165^\circ$. While the
direct CP-violating asymmetry amplitudes of decays $B\to K^{*}_0(1430)\rho(\omega)$ are not large in both scenarios and most of them are less than
$20\%$.
\end{itemize}

\section*{Acknowledgment}
This work is partly supported by Foundation of Henan University of Technology under Grant No.150374.

\end{document}